\documentclass[
    ,final            
  ]
  {aipproc}

\layoutstyle{6x9}

\begin{document}

\title{Low-$Q^2$ partons in p-p and Au-Au collisions
\thanks{Prepared for the Proceedings of the XXXV International Symposium on Multiparticle Dynamics, Krom\^{e}\^{r}\'{i}\^{z}, Czech Republic, August 9-15, 2005.}
}


\author{Thomas A. Trainor\\
\vskip .05in  {\small (STAR Collaboration)}}{
  address={CENPA 354290, University of Washington, Seattle, WA  98195 }
}

\begin{abstract}

We describe correlations of low-$Q^2$ parton fragments on transverse rapidity $y_t$ and angles $(\eta,\phi)$ from p-p and Au-Au collisions at $\sqrt{s} =$ 130 and 200 GeV. Evolution of correlations on $y_t$ from p-p to more-central Au-Au collisions shows evidence for parton dissipation. Cuts on $y_t$ isolate angular correlations on $(\eta,\phi)$ for low-$Q^2$ partons which reveal a large asymmetry about the jet thrust axis in p-p collisions favoring the azimuth direction. Evolution of angular correlations with increasing Au-Au centrality reveals a rotation of the asymmetry to favor pseudorapidity. Angular correlations of transverse momentum $p_t$ in Au-Au collisions access temperature/velocity structure resulting from low-$Q^2$ parton scattering. $p_t$ autocorrelations on $(\eta,\phi)$, obtained from the scale dependence of $\langle p_t \rangle$ fluctuations, reveal a complex parton dissipation process in heavy ion collisions which includes the possibility of collective bulk-medium recoil in response to parton stopping.

\end{abstract}

\maketitle

\section{Introduction}

QCD theory predicts that many low-$Q^2$ scattered gluons (minijets) should be produced in relativistic nuclear collisions at RHIC energies, with rapid parton thermalization as the source of the colored medium (quark-gluon plasma) in heavy ion collisions~\cite{QCD,theor0,theor1,theor2}. If so, we should discover evidence of partons with $Q/2 =$ 1 $-$ 5 GeV in the correlation structure of final-state hadrons. The discovery and analysis of nonperturbative low-$Q^2$ parton fragment correlations has motivated development of novel techniques, including the use of transverse rapidity $y_t$ rather than momentum $p_t$ and formation of angular {\em joint autocorrelations}. Fluctuations of event-wise mean $p_t$ $\langle p_t \rangle$ \cite{Phenix,ptprc} may isolate fragments from low-$Q^2$ partons and determine the properties of the corresponding medium. A measurement of excess $\langle p_t \rangle$ fluctuations in Au-Au collisions at 130 GeV revealed a large excess of fluctuations compared to independent-particle $p_t$ production~\cite{ptprc}. 

Angular correlations of fragments from hard-scattered partons (jets) were first observed on $(\eta,\phi)$ at large $p_t$ and with increasing $\sqrt{s}$~\cite{jets}. In a conventional high-$p_t$ study of parton fragmentation to a hadron jet the relevant issues are the fragment distribution on $p_t$ and the angular distribution, both relative to the parton momentum (jet thrust axis). In contrast, we adopt no {\em a priori} jet or factorization hypothesis. We study minimum-bias two-particle distributions on transverse rapidity space $(y_{t1},y_{t2})$ to obtain fragment {\em distributions} (not fragmentation {\em functions}) and on angle space $(\eta_1,\eta_2,\phi_1,\phi_2)$ to obtain fragment angular correlations. Particle pairs are treated symmetrically, as opposed to asymmetric `trigger' and `associated' particle combinations. We observe that correlations obtained with this minimum-bias analysis, in contrast to the conventional trigger-particle approach, represent the {\em majority} of parton fragment pairs in nuclear collisions, those with $p_{t1} \sim p_{t2} \sim 1$ GeV/c. While we attempt to understand QCD in A-A collisions we should also revisit its manifestations in elementary collisions, where novel phenomena are still emerging.



\section{p-p Initial Survey and Analysis Method}

\begin{figure}[h]
\begin{minipage}{2.9in}
\begin{center}
 \includegraphics[width=1.4in,height=1.4in]{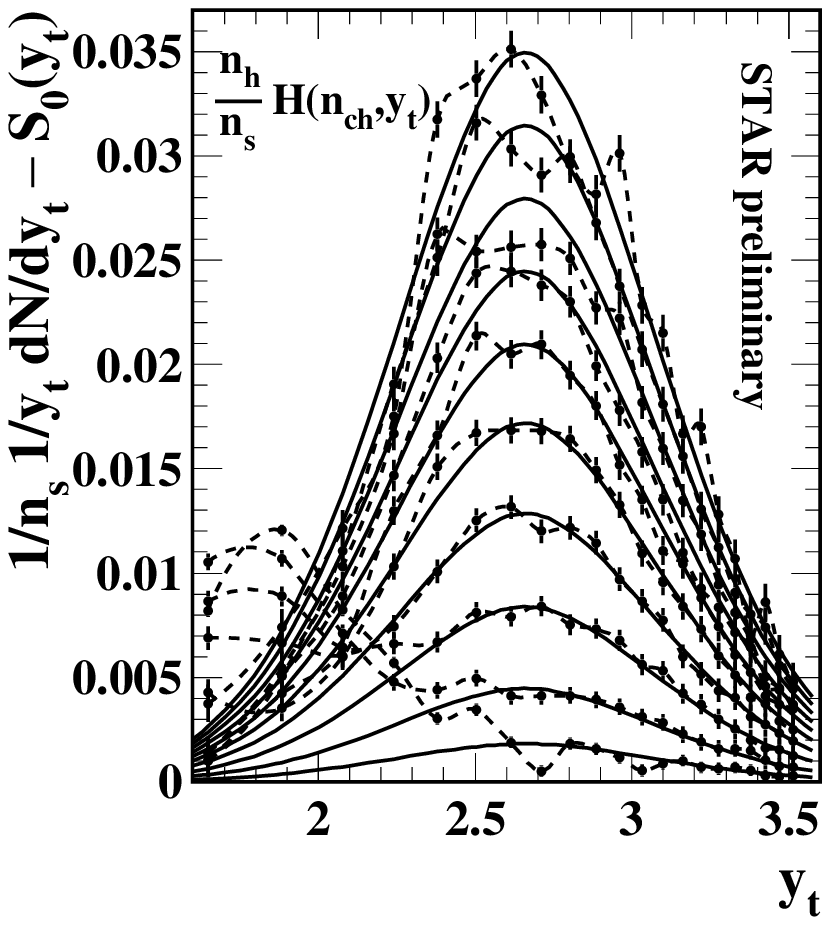} 
\includegraphics[width=1.4in,height=1.4in]{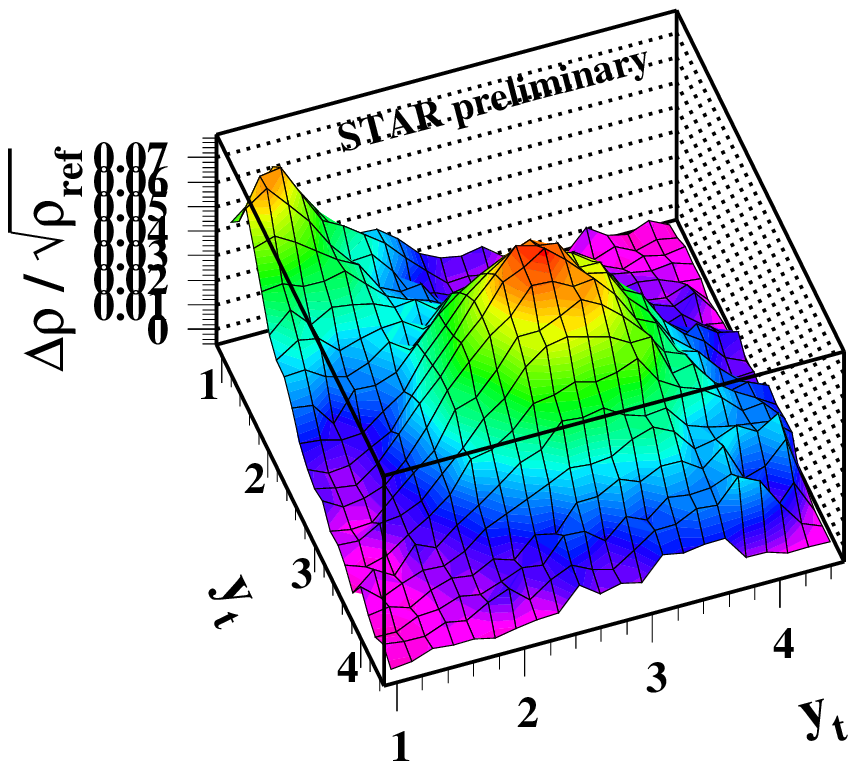}
\end{center} 
\end{minipage}
\hfil
\begin{minipage}{2.9in}
\begin{center}
 \includegraphics[width=1.4in,height=1.4in]{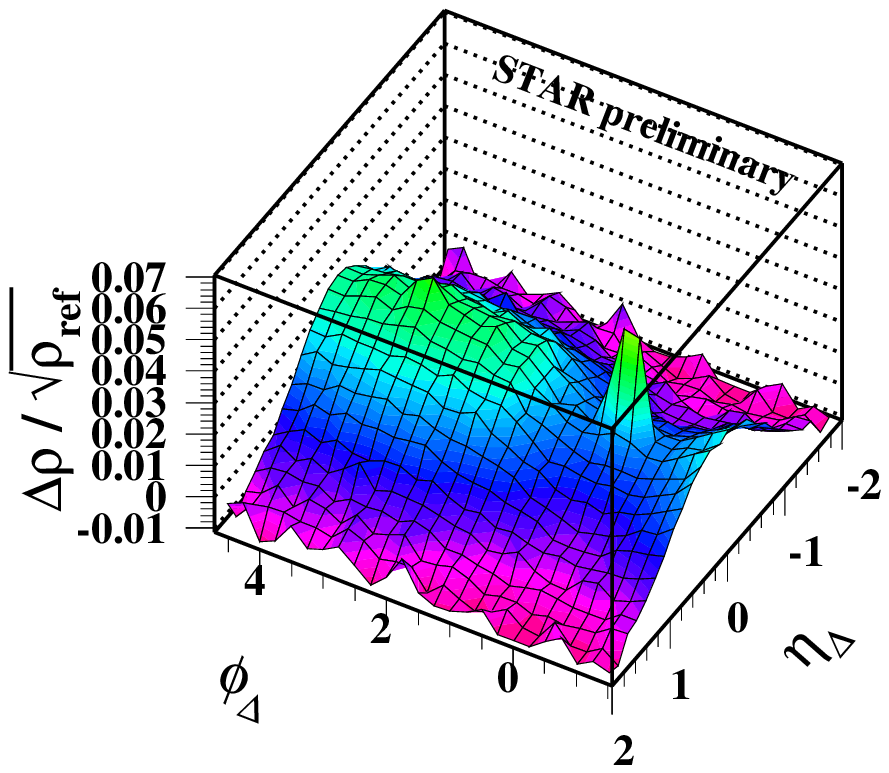} 
\includegraphics[width=1.4in,height=1.4in]{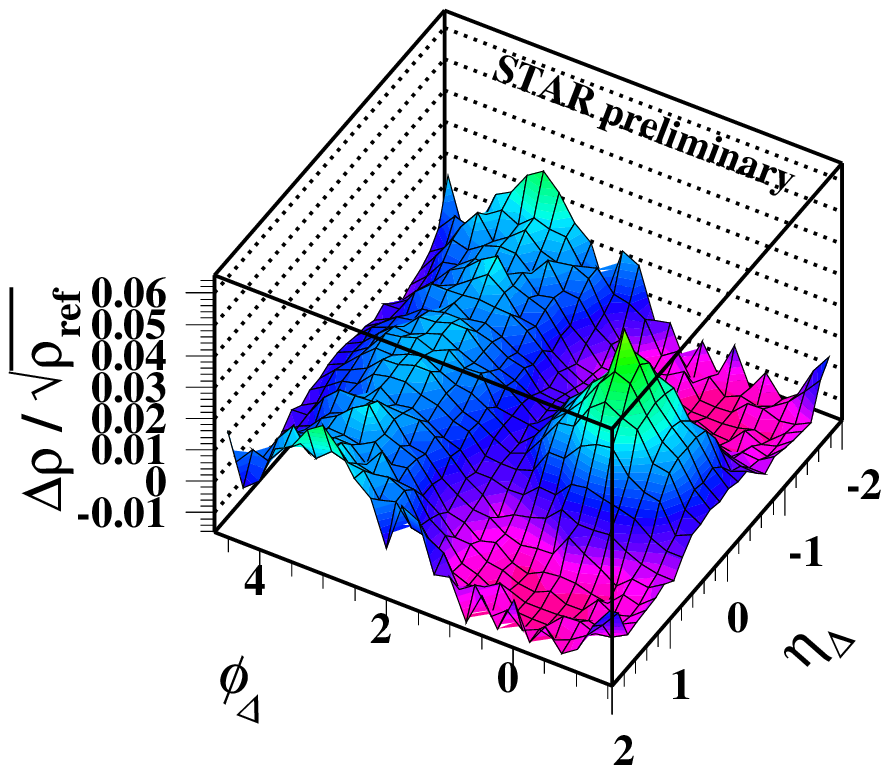} 
\end{center}
\end{minipage}\hspace{0pc}%
\caption{\label{fig1} Hard components of $p_t$ spectra {\em vs} $n_{ch}$ plotted on transverse rapidity $y_t$; low-$Q^2$ parton and string fragment distributions on $(y_{t1},y_{t2})$ and angular difference variables $(\eta_\Delta,\phi_\Delta)$, all for p-p collisions.}
\end{figure}

The reference system for low-$Q^2$ partons in A-A collisions is the {\em hard component} of correlations in p-p collisions. The single-particle $p_t$ spectrum for p-p collisions at $\sqrt{s} = 200$ GeV can be decomposed into soft and hard components on the basis of event multiplicity~\cite{jeffismd}. When a fixed L\'evy distribution~\cite{levy} is subtracted from spectra for ten multiplicity $n_{ch}$ classes we obtain distributions in Fig.~\ref{fig1} (first panel), described by a gaussian shape (solid curves) independent of $n_{ch}$. Transverse rapidity is defined by $y_t \equiv \ln\{(m_t + p_t)/m_0\}$, with $m_0$ assigned as the pion mass. That novel result motivated a study of two-particle correlations on $y_t \times y_t$~\cite{jeffismd}. The minimum-bias distribution in Fig.~\ref{fig1} (second panel) exhibits separate soft and hard components interpreted as longitudinal string fragments (smaller $y_t$) and transverse parton fragments (larger $y_t$), the latter corresponding to the gaussians in the first panel. 

Soft and hard components on $y_t$ produce corresponding structures in joint angular autocorrelations on $(\eta_\Delta,\phi_\Delta)$. In the third panel, string-fragment (soft) correlations for unlike-sign pairs are determined by local charge and transverse-momentum conservation (the sharp peak at the origin is conversion electrons). Minimum-bias parton fragments (hard) in the fourth panel produce classic jet correlations, with a {\em same-side} ($\phi_\Delta < \pi/2$) jet cone at the origin and an {\em away-side} ($\phi_\Delta > \pi/2$) ridge corresponding to the broad distribution of parton-pair centers of momentum. 
These angular {\em autocorrelations} do not rely on a leading or trigger particle and provide unprecedented access to low-$Q^2$ partons. 

Density ratio $\Delta \rho / \sqrt{\rho_{ref}}$ is related to Pearson's  {\em correlation coefficient}~\cite{pearson}. For event-wise particle counts $n_a$ and $n_b$ in histogram bins $a$ and $b$ on space $x$ Pearson's coefficient is $ r_{ab} \equiv \overline{(n - \bar n)_a(n - \bar n)_b} / \sqrt{\overline{(n - \bar n)_a^2}\, \,\overline{(n - \bar n)_b^2}}$. That suggests the form of density ratio $\Delta \rho / \sqrt{\rho_{ref}} \equiv 1/\epsilon_x \,\, \overline{(n - \bar n)_a(n - \bar n)_{b}} / \sqrt{\bar n_a\, \bar n_{b}}$, where $\epsilon_x$ is the histogram bin size on $x$ and Poisson values of the variances in the denominator have been substituted. $\Delta \rho / \sqrt{\rho_{ref}}$ measures correlated pairs {\em per particle} on $(y_{t1},y_{t2})$, $(\eta_{1},\eta_{2})$ and $(\phi_{1},\phi_{2})$. For angular correlations we combine spaces $(\eta_{1},\eta_{2})$ and $(\phi_{1},\phi_{2})$ in a {\em joint autocorrelation}, providing projection {\em by averaging} to a lower-dimensional space with little information loss. An autocorrelation with index $k$ on primary space $x$ is obtained by averaging correlation coefficients $r_{a,a+k}$ over index $a$ along each $k^{th}$ diagonal on $(x_1,x_2)$. For space $(\eta,\phi)$ we average simultaneously along diagonals on $(\eta_{1},\eta_{2})$ and $(\phi_{1},\phi_{2})$ to obtain a {\em joint} autocorrelation on angular {\em difference variables} $\eta_\Delta \equiv \eta_1 - \eta_2$ and $\phi_\Delta \equiv \phi_1 - \phi_2$.

\section{Number Correlations on $y_t \times y_t$}

\begin{figure}[h]
\begin{minipage}{2.9in}
\begin{center}
 \includegraphics[width=1.4in,height=1.4in]{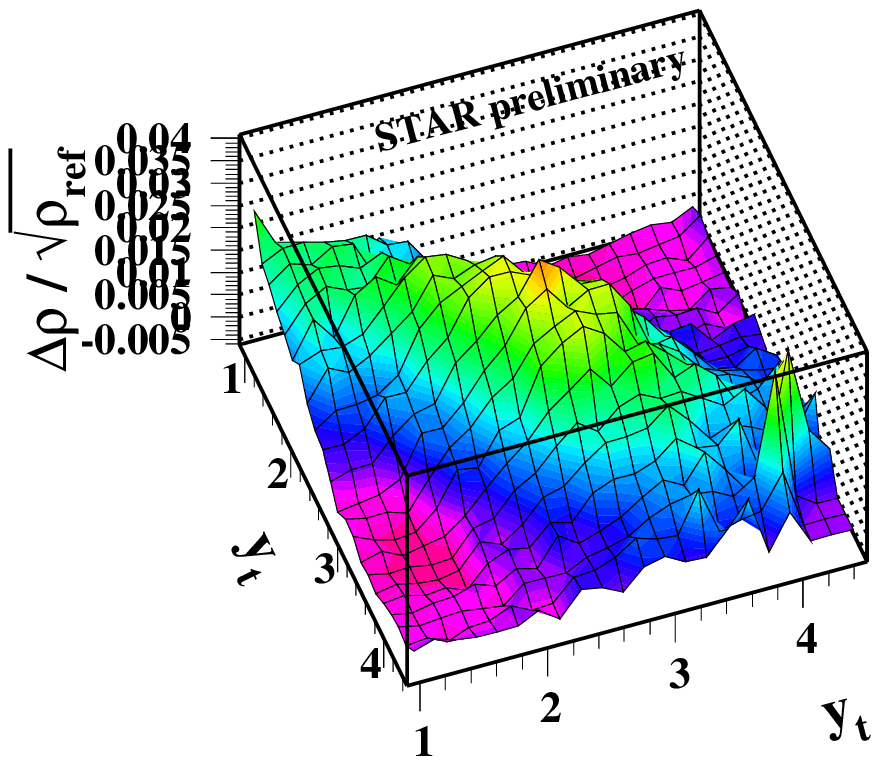} 
\includegraphics[width=1.4in,height=1.4in]{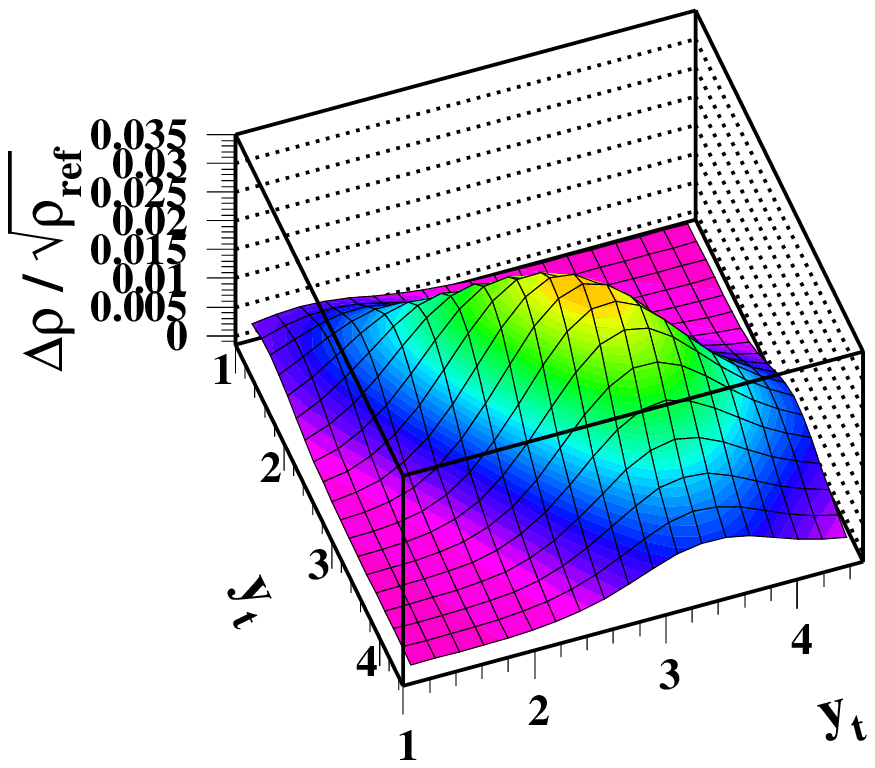}
\end{center} 
\end{minipage}
\hfil
\begin{minipage}{2.9in}
\begin{center}
 \includegraphics[width=1.4in,height=1.4in]{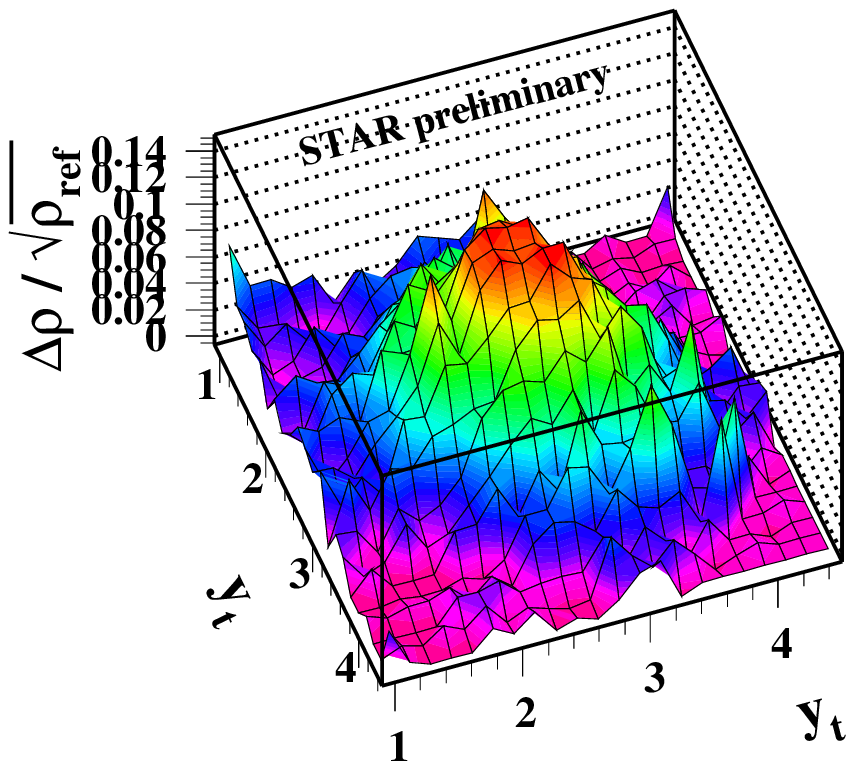} 
\includegraphics[width=1.4in,height=1.4in]{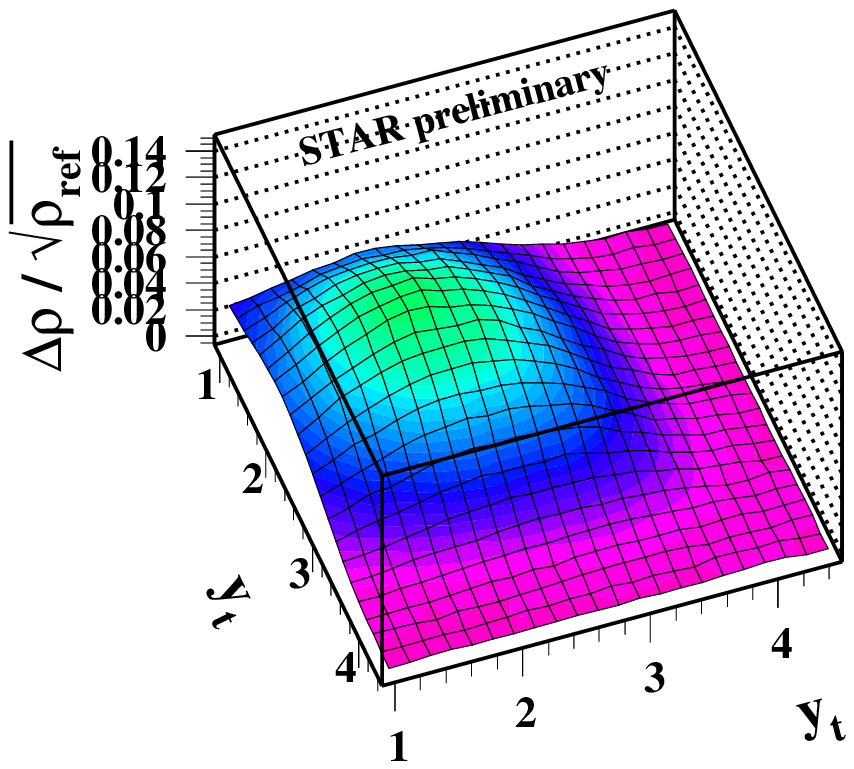} 
\end{center}
\end{minipage}\hspace{0pc}%
\caption{\label{fig2} Parton fragment distributions on $(y_{t1},y_{t2})$ for p-p data (US pairs), a model based on measured fragmentation functions, Au-Au mid-perpheral and Au-Au mid-central, all at $\sqrt{s_{NN}} = 200$ GeV.}
\end{figure}

Particle pairs from nuclear collisions can be separated on difference variable $\phi_\Delta$ into  {same-side} (SS) and {away-side} (AS) pairs. Fig.~\ref{fig2} (first panel) shows unlike-sign (US) SS pairs on $(y_t,y_t)$. 
The US hard component is a peak at $y_t \sim 2.8$ ($p_t \sim$ 1 GeV/c) elongated along $y_{t\Sigma} \equiv y_{t1} + y_{t2}$ and running into the soft (string fragment) component at small $y_t$. 

The same-side unlike-sign (US) $(y_t,y_t)$ correlations in Fig.~\ref{fig1} (first panel) can be interpreted as a two-particle {\em intra}-jet fragment distribution which we now model: we combine single-particle fragmentation functions~\cite{frag} with expectations for two-particle correlations to sketch the parameterization of a two-particle fragment distribution shown in the second panel. We observe that single-particle fragmentation functions plotted on transverse rapidity $y_t$ have a simple form represented by a beta distribution. We therefore construct a parton-fragment joint distribution on $(y_t,y_t)$. By symmetrizing that distribution to approximate a fragment-fragment joint distribution we obtain the intra-jet fragment correlations shown in Fig.~\ref{fig2} (second panel). For low-$Q^2$ partons, which dominate the minimum-bias parton distribution, the two-particle fragment distribution is symmetric about the $y_{t\Sigma}$ diagonal. At larger $y_{t\Sigma}$ the distribution bifurcates symmetrically, the equivalent branches representing a continuum of conventional pQCD {\em conditional} fragmentation functions on fragment $y_t$ {\em given} parton $y_t$.

The right two panels represent evolution with Au-Au centrality. For mid-peripheral collisions (third panel) the string fragments at smaller $y_t$ are eliminated~\cite{axialci} and there is already some slight attenuation at larger $y_t$ due to parton dissipation. For mid-central collisions (fourth panel) the fragment distribution is transported {\em en mass} to lower $y_t$, approaching the limiting case of random temperature variation and hydrodynamics~\cite{mtxmt}. The reduction with centrality at larger $y_t$ corresponds to the variation of $R_{AA}$ on $p_t$~\cite{raapaper}.
 

\section{Number Correlations on $\eta \times \phi$}

\begin{figure}[h]
\begin{minipage}{2.9in}
\begin{center}
 \includegraphics[width=1.4in,height=1.4in]{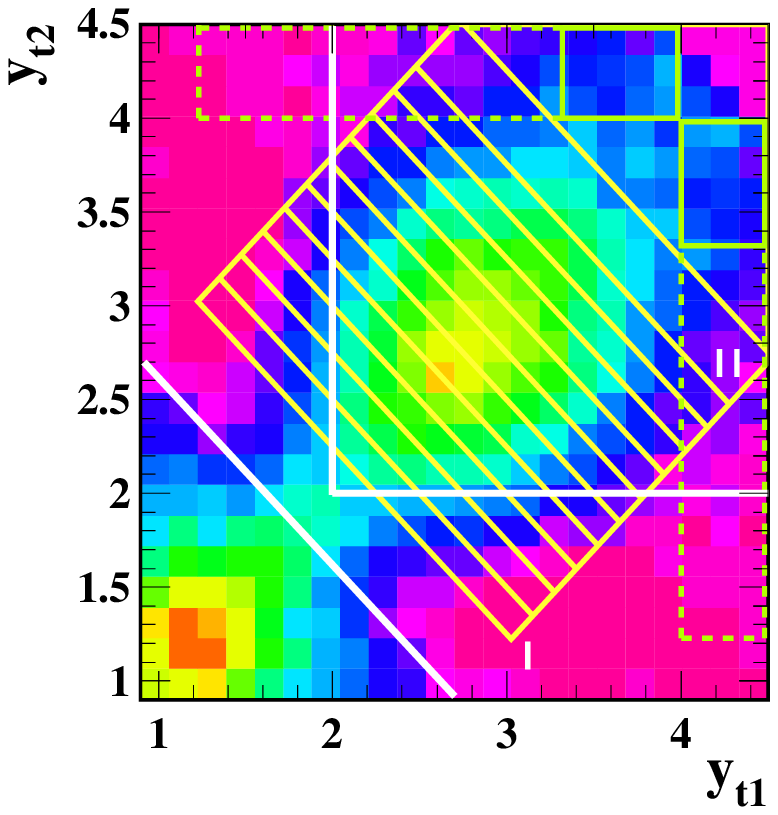} 
\includegraphics[width=1.4in,height=1.4in]{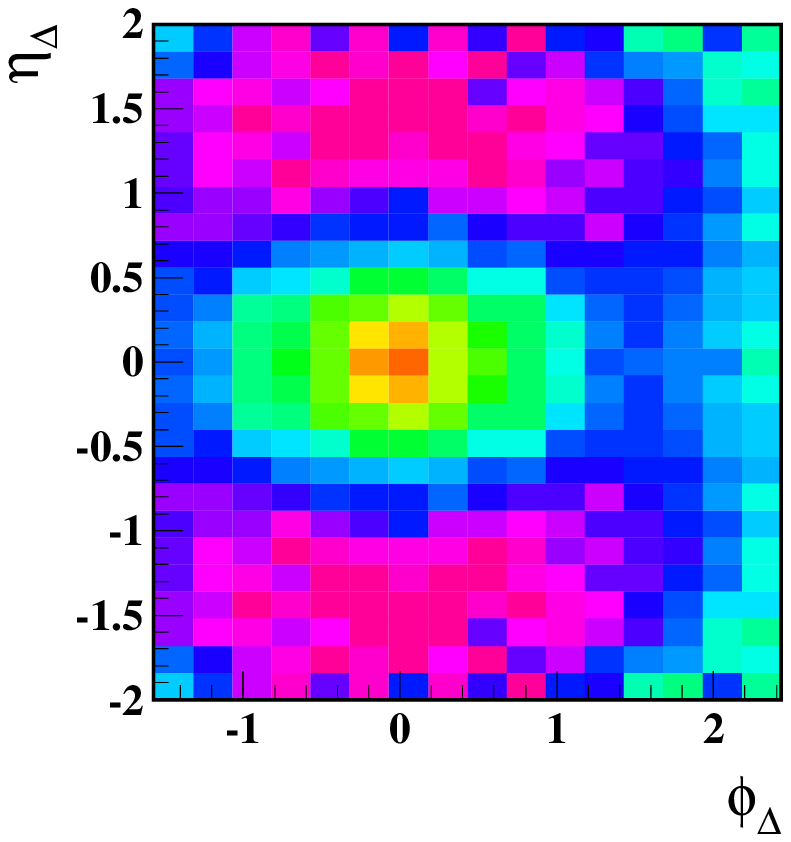}
\end{center} 
\end{minipage}
\hfil
\begin{minipage}{2.9in}
\begin{center}
 \includegraphics[width=1.4in,height=1.4in]{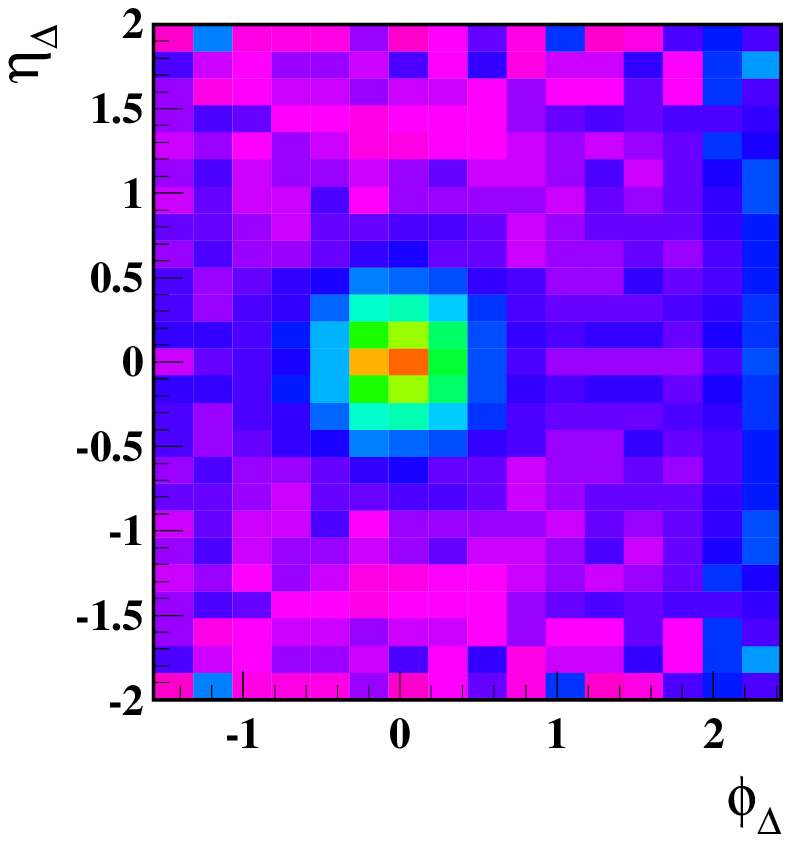} 
\includegraphics[width=1.4in,height=1.4in]{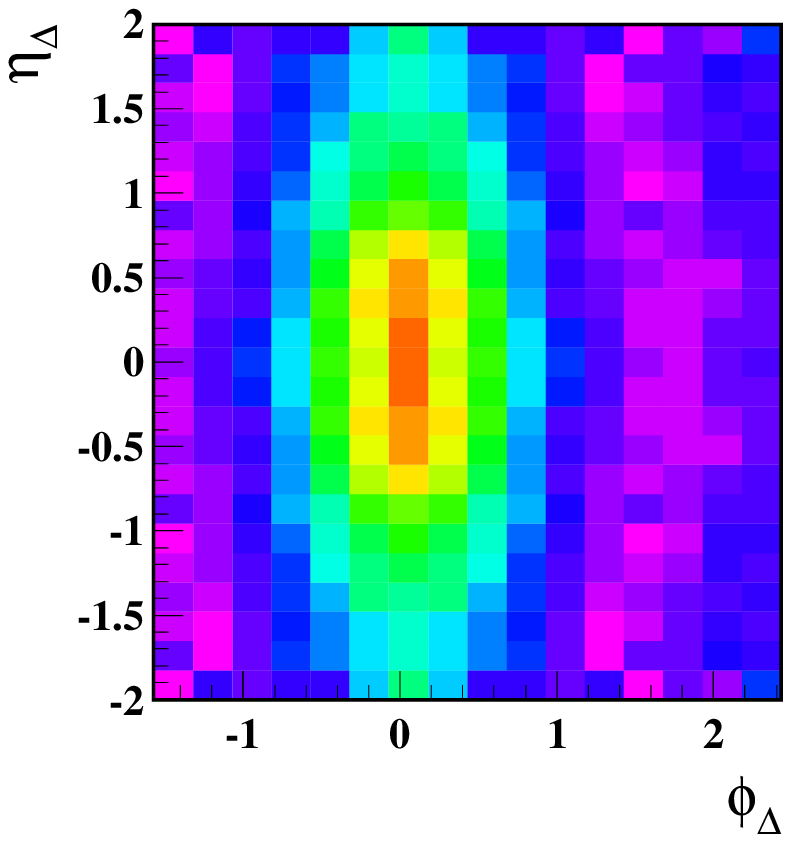} 
\end{center}
\end{minipage}\hspace{0pc}%
\caption{\label{fig3} Binned $(y_{t1},y_{t2})$ cut space, corresponding small-$y_{t\Sigma}$ (bin 2) and large-$y_{t\Sigma}$ (bin 11) angular correlations on $(\eta_\Delta,\phi_\Delta)$ for p-p collisions, and angular correlations for mid-central Au-Au collisions.}
\end{figure}

Jet structure is also characterized by two-particle angular correlations of hadron fragments on $(\eta,\phi)$ complementary to correlations on $(y_{t1},y_{t2})$ described in the previous section. 
The conventional method to describe angular correlations of fragments in the absence of full jet reconstruction is as a {\em conditional} distribution relative to a trigger-particle momentum estimating the parton momentum. In this analysis we use no trigger condition. $(y_{t1},y_{t2})$ correlations provide a cut space for study of angular autocorrelations. Fig.~\ref{fig3} (first panel) shows two-particle correlations for p-p collisions on transverse-rapidity space $(y_{t1},y_{t2})$ binned on sum and difference variables $y_{t\Sigma}$ and $y_{t\Delta} \equiv y_{t1} - y_{t2}$. 

Hard-component fractions for angular correlation measurements are defined by the grid of bins along $y_{t\Sigma}$ (the narrow yellow lines) numbered $1, \cdots, 12$. The solid green boxes in the upper-right corner represent regions explored in {\em leading-particle} analyses based on a high-$p_t$ `trigger' particle~\cite{leading}. The dashed extensions represent cuts for extended associated-particle conditions applied to heavy ion collisions~\cite{fuqiang}. Hard-component correlations on $(\eta_\Delta,\phi_\Delta)$ consist of a same-side peak at the origin and an away-side ridge. The US same-side peak represents intrajet angular correlations of parton fragments (jet cone). These hard-component $(\eta,\phi)$ systematics, fully consistent with conventional expectations for high-$p_t$ jet angular correlations, are observed in this study for pairs of particles with {\em both} $p_t$s as low as 0.35 GeV/c ($y_t \sim 1.6$), {\em much lower than previously observed with leading-particle methods}.

The second and third panels of Fig.~\ref{fig3} show angular autocorrelations
(these plots are 1:1 aspect ratio, hence exclude most or all of the away-side ridge) for bins 2 and 11 on $y_{t\Sigma}$ in Fig.~\ref{fig3} (first panel). In the second panel (bin 2) the most probable combination is two particles each with $p_t \sim 0.6$ GeV/c. The same-side peak (jet cone) is strongly elongated in the azimuth direction, suggesting that some aspect of the parton collision geometry is retained in these soft collisions. The third panel (bin 11) shows correlations corresponding to $p_t \sim 2.5$ GeV/c for each particle. The same-side cone is much narrower and {\em nearly} symmetric, more typical of a high-$p_t$ trigger-particle analysis described by pQCD. The general trend is monotonic reduction of peak widths on  $(\eta_\Delta,\phi_\Delta)$ with increasing $y_{t\Sigma}$ ($Q^2$). The fourth panel shows mid-central Au-Au collisions at $\sqrt{s_{NN}} = 130$ GeV. The near-side peak is strongly elongated on pseudorapidity rather than azimuth, suggesting strong coupling of the fragmenting parton to longitudinal Hubble expansion of the QCD medium~\cite{axialci}.

\section{$\langle p_t \rangle$ Fluctuations and $p_t$ Correlations on $\eta \times \phi$}

\begin{figure}[h]
\begin{minipage}{3.15in}
\begin{center}
 \includegraphics[width=1.55in,height=1.4in]{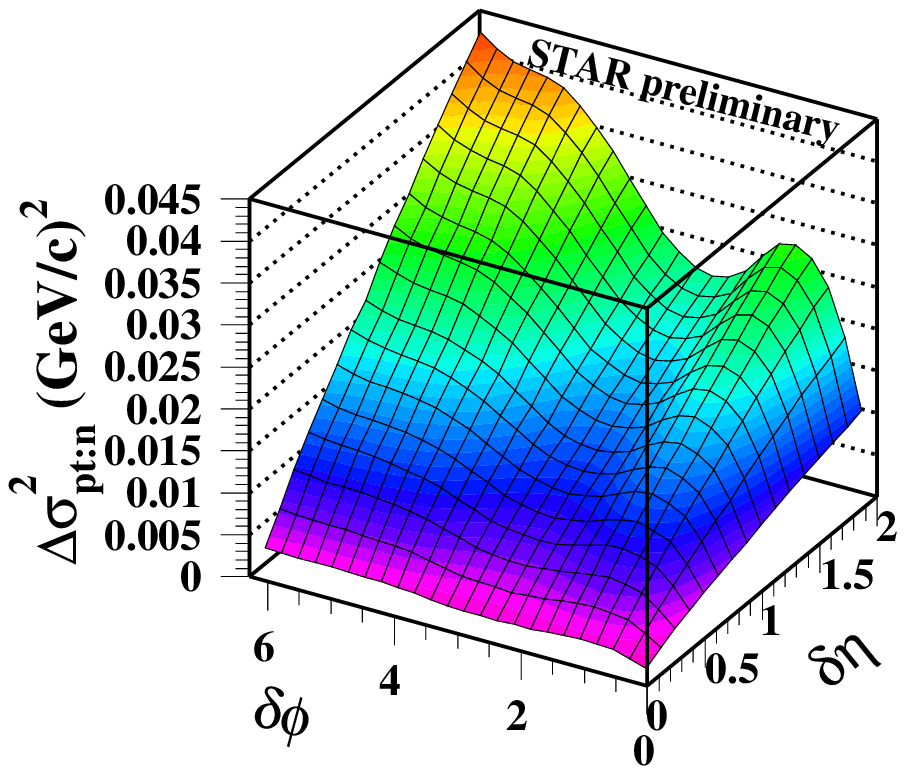} 
\includegraphics[width=1.55in,height=1.4in]{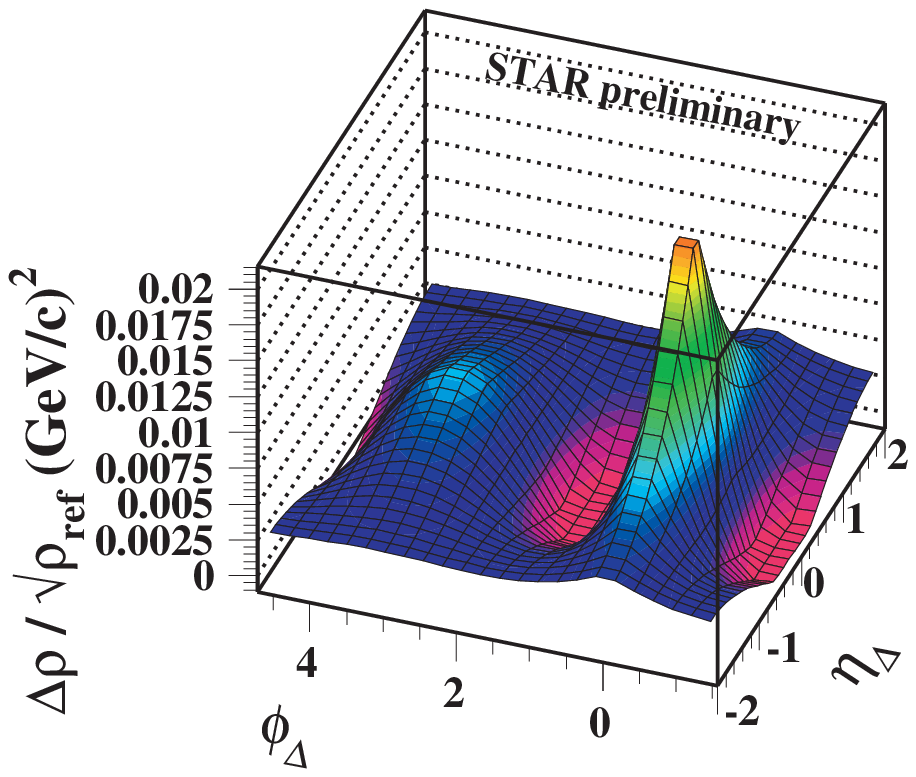}
\end{center} 
\end{minipage}
\hfil
\begin{minipage}{2.6in}
\begin{center}
 \includegraphics[width=1.4in,height=1.in,angle=90]{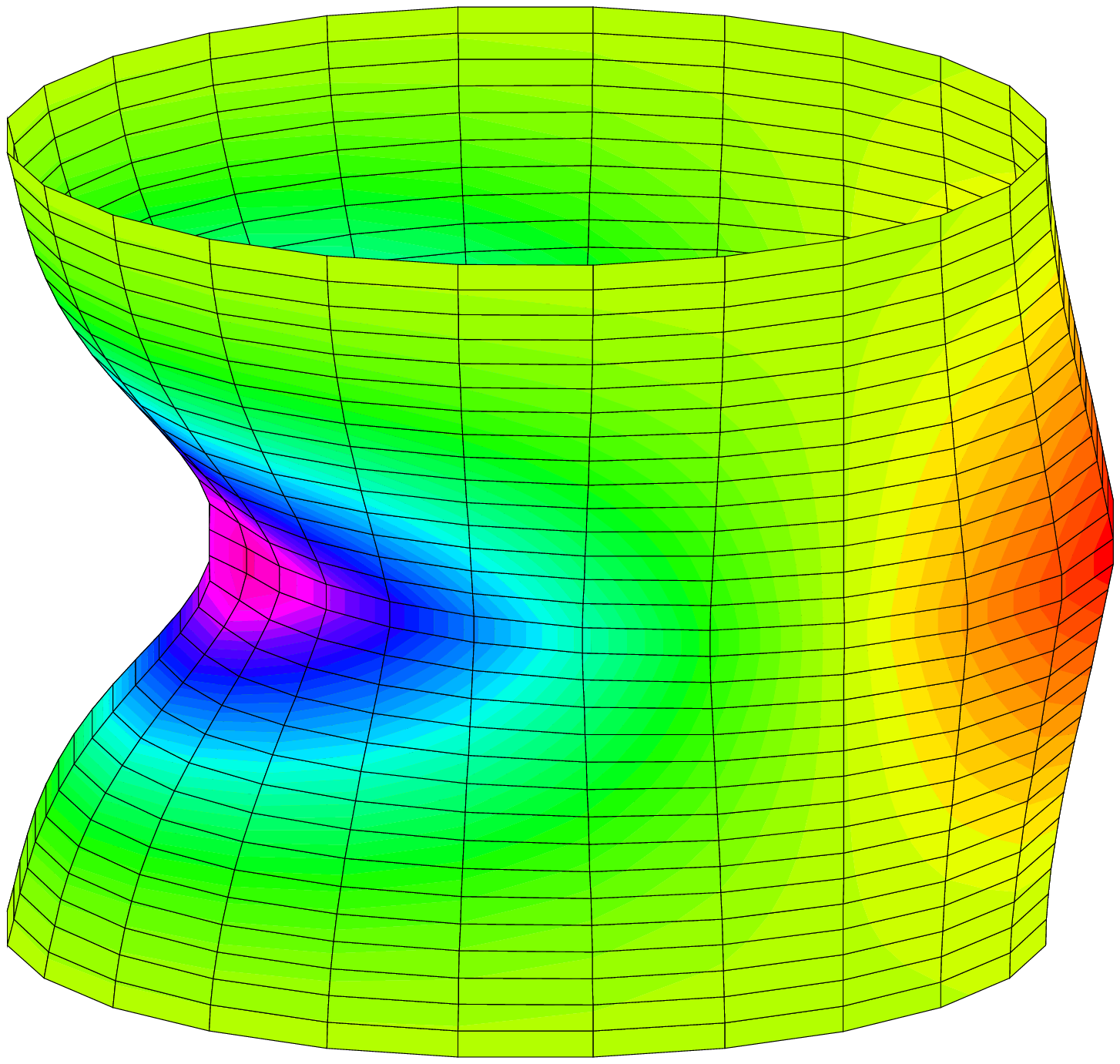} 
\hskip .1in
\includegraphics[width=1.4in,height=1.4in]{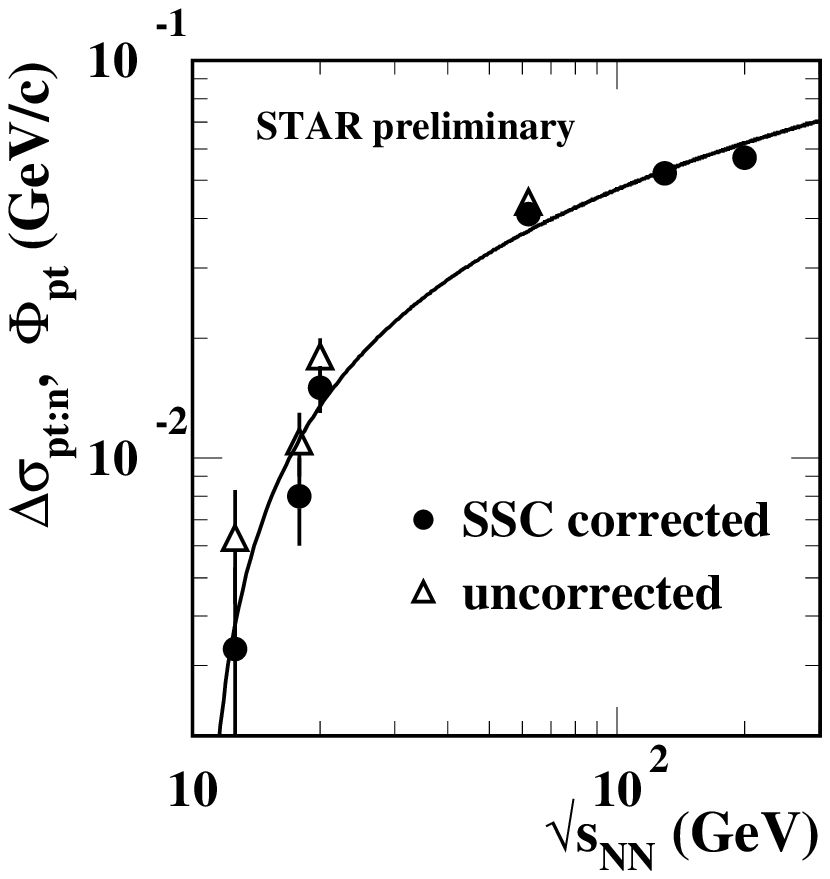} 
\end{center}
\end{minipage}\hspace{0pc}%
\caption{\label{fig4} $\langle p_t \rangle$ fluctuation scale dependence for Au-Au collisions at $\sqrt{s_{NN}} = 200$ GeV, corresponding $p_t$ autocorrelation, same with positive peak subtracted and energy dependence of $\langle p_t \rangle$ fluctuations.}
\end{figure}

By measuring $\langle p_t \rangle$ fluctuation magnitudes as a function of bin size one can recover those aspects of the two-particle $p_t$ correlation structure which depend on the separation of pairs of points, not on their absolute positions. Fig.~\ref{fig4} (first panel) shows fluctuation scale dependence on bin sizes $(\delta \eta,\delta \phi)$ for 20-30\% central Au-Au collisions at 200 GeV~\cite{ptscale}. Fluctuation measurements at the full STAR acceptance~\cite{ptprc} correspond to the single point at the apex of the distribution on scale. By inverting $\langle p_t \rangle$ fluctuation scale dependence~\cite{inverse}, parton fragment distributions are visualized as temperature/velocity structures: joint $p_t$ autocorrelations on angular difference variables $(\eta_\Delta,\phi_\Delta)$. 

Fig.~\ref{fig4} (second panel) shows the corresponding $p_t$ autocorrelation~\cite{ptscale}. We have subtracted azimuth sinusoids independent of pseudorapidity ({\em e.g.,} elliptic flow observed for the first time as a velocity structure), revealing peak structures associated with parton scattering and fragmentation. A three-peak model of that distribution, including separate positive and negative same-side ($\phi_\Delta < \pi / 2$) peaks (the two peaks have very different shapes), provides an excellent fit, with residuals at the percent level. Fig.~\ref{fig4} (third panel) shows the result of subtracting the positive same-side model peak (representing parton fragments) from data in the second panel and plotting the difference with a cylindrical format.  The negative same-side peak can be interpreted as a systematic red shift of local $p_t$ distributions {\em in the neighborhood of} the positive parton fragment peak. The red shift can in turn be interpreted as the result of recoil of the bulk medium in response to stopping the parton {\em partner} of the observed parton (positive same-side peak). 
This detailed picture of parton dissipation, stopping and fragmentation is accessed for the first time with $p_t$ autocorrelations.

Fig.~\ref{fig4} (fourth panel) shows the energy dependence of full-acceptance $\langle p_t \rangle$ fluctuations for central heavy ion collisions at RHIC and SPS measurements by the CERES collaboration (lowest two energies)~\cite{ceres}.  Fluctuation measure $\Delta \sigma_{p_t:n}$ is related to the variance difference by  $\Delta \sigma^2_{p_t:n} \equiv 2 \sigma_{\hat p_t} \, \Delta \sigma_{p_t:n}$, with $\sigma_{\hat p_t}$ the single-particle variance. To good approximation $\Delta \sigma_{p_t:n} \simeq \Phi_{p_t}$, the latter used for the CERES fluctuation measurements. For either measure we observe a dramatic increase in $\langle p_t \rangle$ fluctuations from SPS to RHIC energies where we have demonstrated that $\langle p_t \rangle$ fluctuations are dominated by fragments from low-$Q^2$ parton collisions. 
We observe that $\langle p_t \rangle$ fluctuations vary almost linearly with $\log\{\sqrt{s_{NN}}/10\}$ (solid curve in that panel), suggesting a threshold for {\em observable} parton scattering and fragmentation near 10 GeV.

\section{Summary}

We have presented a survey of two-particle correlations from p-p and Au-Au collisions at RHIC. Correlations from longitudinal string fragmentation and transverse scattered parton fragmentation are clearly distinguished. The jet morphology of low-$Q^2$ partons requires a more general treatment of fragment $p_t$ distributions and angular correlations. Conventional asymmetric treatments of parton fragments in terms of trigger and associated particles cannot access the low-$Q^2$ partons of greatest interest to us. 

Using newly-devised analysis techniques we find that parton fragments are accessible down to hadron $p_t$ = 0.35 GeV/c for both hadrons of a correlated pair. Fragment distributions on transverse rapidity $y_t$ are, for p-p collisions, consistent with measured fragmentation functions in elementary collisions but reveal increasing parton dissipation with greater A-A centrality. Jet angular correlations in p-p collisions show a dramatic asymmetry about the thrust axis at low $Q^2$, with larger width in the azimuth direction possibly related to {\em nonperturbative} details of semi-hard parton collisions, rotating to elongation on pseudorapidity for central Au-Au collisions. 


Inversion of the scale dependence of $\langle p_t \rangle$ fluctuations provides access to $p_t$ angular autocorrelations, revealing a complex parton dissipation process in A-A collisions and possible evidence for bulk-medium recoil in response to parton stopping. We also observe strong energy dependence of $\langle p_t \rangle$ fluctuations, consistent with the dominant role of scattered partons in those fluctuations. Low-$Q^2$ partons, accessed here for the first time by novel analysis techniques including joint autocorrelations, serve as {\em Brownian probes} of the A-A medium at RHIC, being the softest {\em detectable} objects which experience QCD interactions as color charges.


\begin{thebibliography}{9}

\bibitem{QCD} J.~C. Collins and M. Perry {\em Phys. Rev. Lett.}, {\bf 34}, 1353 (1975);\\
	B. Freedman and L. McLerran  {\em Phys. Rev.} D, {\bf 17}, 1109 (1978);\\
Proceedings of Quark Matter 2004, {\em J. Phys.} G, {\bf 30}, various articles.

\bibitem{theor0}
K. Kajantie, P.~V. Landshoff and J. Lindfors
{\em Phys. Rev. Lett.},  {\bf 59}, 2527 (1987).

\bibitem{theor1} 
A.~H. Mueller
{\em Nucl. Phys.} B, {\bf 572}, 227 (2000).

\bibitem{theor2} 
G.~C. Nayak, A. Dumitru, L.~D. McLerran and W. Greiner
{\em Nucl. Phys.} A, {\bf 687}, 457 (2001).

\bibitem{Phenix} K. Adcox et al. (PHENIX Collaboration) {\em Phys. Rev.} C, {\bf 66}, 024901 (2002).

\bibitem{ptprc} J. Adams  et al. (STAR Collaboration) {\em Phys. Rev.} C, {\bf 71}, 064906 (2005). 
 
\bibitem{jets} A.~L.~S. Angelis {et al.} (CCOR) {\em Phys. Lett.} B, {\bf 97}, 163 (1980);\\
 C. Albajar {et al.} (UA1)
{\em Nucl. Phys.} B, {\bf 309}, 405 (1988).

\bibitem{jeffismd}  R.~J. Porter and T.~A. Trainor  (STAR Collaboration)
  {\em Acta Phys. Polon.} B, {\bf 36}, 353 (2005).

\bibitem{levy}  G. Wilk and Z. Wlodarczyk 
{\em Phys. Rev. Lett.}, {\bf 84}, 2770 (2000).

\bibitem{pearson} K. Pearson {\em Phil. Trans. Royal Soc.}, {\bf 187}, 253 (1896).

\bibitem{frag} M.~Z. Akrawy {et al.}  (OPAL Collaboration)
  {\em Phys. Lett.} B, {\bf 247}, 617 (1990).

\bibitem{axialci} J. Adams  et al. (STAR Collaboration)  nucl-ex/0411003.

\bibitem{mtxmt} J. Adams  et al. (STAR Collaboration)  nucl-ex/0408012.

\bibitem{raapaper} 
 C. Adler {et al.} (STAR Collaboration)
  {\em Phys. Rev. Lett.},  {\bf 89}, 202301 (2002).

\bibitem{leading} C. Adler {et al.} (STAR Collaboration) {\em Phys. Rev. Lett.}, {\bf 90}, 082302 (2003).

\bibitem{fuqiang} J. Adams  et al. (STAR Collaboration)  nucl-ex/0501016.

\bibitem{ptscale} J. Adams  et al. (STAR Collaboration)  nucl-ex/0509030.

\bibitem{inverse} T.~A. Trainor, R.~J. Porter and D.~J. Prindle {\em J. Phys.} G,  {\bf 31}, 809 (2005). 

\bibitem{ceres}  D. Adamova {et al.}  (CERES Collaboration)
   {\em Nucl. Phys.} A, {\bf 727}, 97 (2003).

\end{thebibliography}
\end{document}